\documentstyle[12pt,epsfig]{article}

\newcommand{\beq}{\begin{equation}} 
\newcommand{\eeq}{\end{equation}} 
\newcommand{\barr}{\begin{eqnarray}} 
\newcommand{\earr}{\end{eqnarray}} 

\newcommand{\bfk}{\mbox{\boldmath $k$}} 
\newcommand{\bfy}{\mbox{\boldmath $y$}} 
\newcommand{\bfx}{\mbox{\boldmath $x$}} 
\newcommand{\bfz}{\mbox{\boldmath $z$}} 
\newcommand{\pup}{p^\uparrow}

\newcommand{\Lup}{\Lambda^\uparrow} 
 
\newcommand{\simorder}{\raisebox{-4pt}{$\, \stackrel{\textstyle >}{\sim} \,$}} 
\newcommand{\NP}[1]{{\it Nucl.\ Phys.}\ {\bf #1}}

\newcommand{\PL}[1]{{\it Phys.\ Lett.}\ {\bf #1}} 
\newcommand{\PR}[1]{{\it Phys.\ Rev.}\ {\bf #1}} 
\newcommand{\PRLL}[1]{{\it Phys.\ Rev.\ Lett.}\ {\bf #1}} 
\newcommand{\IJMP}[1]{{\it Int.\ J.\ Mod.\ Phys.}\ {\bf #1}} 
\newcommand{\MPL}[1]{{\it Mod.\ Phys.\ Lett.}\ {\bf #1}}

\begin{document}
\begin{flushright} 
INFNCA-TH0022 \\ 
hep-ph/0011390 \\ 
\end{flushright} 
\vskip 1.5cm 
\begin{center} 

{\bf
Phenomenology of single transverse spin asymmetries\footnote{Talk 
delivered by U.\ D'Alesio at ``VIII Convegno su Problemi di Fisica Nucleare 
Teo\-ri\-ca'', October 18-20, 2000, Cortona, Italy.}
}\\
\vskip 0.8cm
{\sf U.\ D'Alesio and F.\ Murgia}
\vskip 0.5cm
Dipartimento di Fisica, Universit\`a di Cagliari and \\ 
INFN, Sezione di Cagliari, 
C.P. 170, I-09042 Monserrato (CA), Italy\\
\end{center} 
\vskip .5cm 
\noindent 
\noindent
\begin{center} 
{\bf Abstract}
\end{center}  
A unified and consistent phenomenological 
approach to single transverse  spin asymmetries in the
framework of perturbative QCD, with the inclusion of  
a new class of spin and
$\bfk_\perp$ dependent distribution and fragmentation
functions, is presented. As an example,  
results on $A_N(\pup p\to \pi X)$ and $P_\Lambda(pp\to\Lup X)$ are shown.

\vspace*{30pt}
\noindent
{\bf Introduction}
\vspace*{10pt}

Perturbative QCD with its factorization theorems 
has been in the last 
decades the main tool to study hard hadronic processes. 
An important feature of  pQCD factorization 
theorems is the use of {\em collinear partonic 
configurations}, 
where collinear means that the transverse momentum ($\bfk_\perp$)
dependence of the parton relative to the parent hadron or of the
hadron relative to the fragmenting parton is integrated out.
Even if this well-established approach is able to explain
many properties of hard processes, 
a class of phenomena involving polarization degrees of
freedom, like single transverse spin asymmetries, 
seem to be out of this understanding.
Indeed according to  collinear factorization theorems the asymmetry 
for the process $A B \to C X$ with $A$ or $C$ 
transversely polarized (relative to the scattering plane), 
defined as
\beq
A_N = \frac{d\sigma^\uparrow - d\sigma^\downarrow}
{d\sigma^\uparrow +d\sigma^\downarrow}
\eeq
($d\sigma^\uparrow \equiv E_Cd^3\sigma^\uparrow/d^3p_C$ 
with $A^\uparrow$ or $C^\uparrow$),   
is almost vanishing. 
This comes unavoidably from the fact that the asymmetry at 
the partonic level is negligible.
On the other hand a lot of data indicate values for such asymmetries  
of the order of 30-40\% in size in some kinematical regions.
In particular we refer to the 
observed large asymmetries in 
inclusive pion production in $p^\uparrow p \to \pi X$ \cite{e704}
and to transverse $\Lambda$ 
polarization in $pp\to\Lambda^\uparrow X$~\cite{data}.

In the last years a new extended formalism based on 
pQCD factorization theorems, including spin and 
intrinsic $\bfk_\perp$  effects,  has been
formulated~\cite{siv}$-$\cite{amcol},   
and some promising phenomenological 
applications have been performed.
In this contribution we will summarize the main features of this
approach which involves a new class of 
partonic distribution and fragmentation
functions. 
We will also present and discuss  
the main results obtained in the understanding
of such single spin asymmetries (SSA) 
and how we can get reasonable information
on these new distribution functions by fits on available data.

\vspace*{30pt}
\noindent
{\bf Formalism}
\vspace*{10pt}

The original suggestion that intrinsic $\bfk_\perp$ of the quarks in
the distribution functions might give origin to single spin
asymmetries was first made by Sivers~\cite{siv}. 
A similar suggestion, this time
for the transverse momentum of the observed hadron relative to the 
fragmenting quark was later formulated by Collins~\cite{col}.

More generally we can define a new class of non perturbative functions
and generalize 
the usual factorization theorem for the analysis of 
$AB\to CX$ processes, at large energies and moderately
large $p_T$, with the inclusion of spin and intrinsic (partonic)
transverse momentum effects.
These new, twist-two, spin and $\bfk_\perp$ dependent
partonic distribution/fragmentation functions originate from soft,
non-perturbative dynamics, which induces correlations between the
intrinsic transverse momentum of, e.g., an unpolarized parton(hadron)
inside (produced in the fragmentation of) a transversely polarized
hadron(parton); this in turn results in an azimuthal asymmetry for the
$\bfk_\perp$ dependence of the parton(hadron) probability
distribution. The same is valid when the transversely polarized
particle is the final parton(hadron).  

In the sequel we  will refer to the parton as 
a quark ($q$) and to the incoming hadron as a proton ($p$).
For spin one-half hadrons we have
\barr
\Delta^N f_{q/p^\uparrow}(x,\bfk_{\perp}) & = &
\hat f_{q/p^\uparrow}(x,\bfk_{\perp}) -
\hat f_{q/p^\downarrow}(x,\bfk_{\perp}) 
\hspace*{12pt} [{\rm Sivers (90)}] \label{def1}\\
\Delta^N f_{q^\uparrow/p}(x,\bfk_{\perp}) & = &
\hat f_{q^\uparrow/p}(x,\bfk_{\perp}) -
\hat f_{q^\downarrow/p}(x,\bfk_{\perp}) \label{def2} \\
\Delta^N D_{h^\uparrow/q}(z,\bfk_{\perp}) & = &
\hat D_{h^\uparrow/q}(z,\bfk_{\perp}) -
\hat D_{h^\downarrow/q}(z,\bfk_{\perp}) \label{ded1} \\
\Delta^N D_{h/q^\uparrow}(z,\bfk_{\perp}) & = &
\hat D_{h/q^\uparrow}(z,\bfk_{\perp}) -
\hat D_{h/q^\downarrow}(z,\bfk_{\perp}) 
 \hspace*{5pt} [{\rm Collins (93)}] \label{ded2}\,.
\earr
Notice that by rotational invariance 
$
\hat f_{q/p^\downarrow}(x,\bfk_{\perp}) = 
\hat f_{q/p^\uparrow}(x,-\bfk_{\perp}) 
$ and so on.

\noindent
The Sivers function, Eq.~(\ref{def1}), is 
the difference between the number density  
$\hat f_{q/p^\uparrow}(x,\bfk_{\perp})$ and 
$\hat f_{q/p^\downarrow}(x,\bfk_{\perp})$ 
of quarks $q$, with all possible polarizations, longitudinal momentum 
fraction $x$ and intrinsic transverse momentum $\bfk_\perp$, inside a
transversely polarized proton with spin $\uparrow$ or $\downarrow$.
The other functions have similar and self-explanatory meaning.
An analogous set of spin and $\bfk_\perp$ dependent functions
is given, 
following the order in Eqq.~(\ref{def1})-(\ref{ded2}),   
by $f_{1T}^\perp$,  $h_1^\perp$,  $D_{1T}^\perp$ and  $H_1^\perp$~
\cite{Mulders-Tangerman-96,dan1,dan2}. 
All these functions  are $\bfk_\perp$-odd  
(they vanish when
$\bfk_\perp \to 0$) and by parity invariance they have to vanish when
the hadron/quark transverse spin has no component perpendicular to
$\bfk_\perp$, so that for instance
\beq
\label{sin}
\Delta^N f_{q/p^\uparrow}(x,\bfk_{\perp}) \sim k_\perp {\rm
sin}\,\alpha 
\eeq  
where $\alpha$ is the angle between $\bfk_\perp$ and the $\uparrow$
direction. 

These functions are also T-odd, that is they would be zero due to time
reversal invariance. The appearance of these 
 distribution/fragmentation functions 
therefore requires that some soft
initial/final state interactions 
are at work. 
However, initial state interactions
might pose severe
problems because they could spoil the factorization itself. 
The same is not true for  final state interactions that can
be at work among the hadron and the remnants of the fragmenting quark.
Another property of these functions is their chirality: the two
functions in Eq.~(\ref{def2}) and Eq.~(\ref{ded2}) (Collins function) 
are chiral-odd, that is they couple quarks with left- and
right-handed chiralities. This implies, since pQCD interactions
conserve chirality, that they must appear together with another
chiral-odd function (see below) or accompanied by a mass term.
The two functions in Eq.~(\ref{def1}) (Sivers function) and Eq.~(\ref{ded1})
are instead chiral-even and appear together with unpolarized
distribution/fragmentation functions. 

In this formalism, the  asymmetry for the 
process $p^\uparrow p \to \pi X$ at leading twist and leading order in
$\bfk_\perp$ can be expressed as  
\barr
\label{pi}
\lefteqn{2d\sigma^{\rm unp}  A_N  = 
	\sum_{abcd}\int \frac{dx_a dx_b}{\pi z}} \\
&  & \times 
	\Big\{  \int d^2\bfk_{\perp a}\,
	\Delta^N f_{a/p^\uparrow}(x_a,\bfk_{\perp a})\, f_{b/p}(x_b)\,
	\frac{d\hat\sigma}{d\hat t} (x_a,x_b;\bfk_{\perp a})\,
	D_{\pi/c}(z) \nonumber\\
&  &   +
	\int d^2\bfk_{\perp c}\, \Delta_T f_{a/p}(x_a)\, f_{b/p}(x_b)\,
	\Delta_{NN}^{(ac)}\hat\sigma(x_a,x_b;\bfk_{\perp c})\,
	{\Delta^N D_{\pi/c^\uparrow}(z,\bfk_{\perp c})} \nonumber\\
&  &   +
	\int d^2\bfk_{\perp b}\, \Delta_T f_{a/p}(x_a)\,
	\Delta^Nf_{b^\uparrow/p}(x_b,\bfk_{\perp b})\,
	\Delta_{NN}^{'(ab)}\hat\sigma(x_a,x_b;\bfk_{\perp b})\, D_{\pi/c}(z) 
	\Big\}\nonumber
\earr
where the elementary partonic interactions are
\beq
\Delta_{NN}^{(ac)}\hat\sigma  \equiv
\frac{d\hat\sigma^{a^\uparrow b\to c^\uparrow d}}{d\hat t} -
\frac{d\hat\sigma^{a^\uparrow b\to c^\downarrow d}}{d\hat t}
\hspace*{20pt}
\Delta_{NN}^{'(ab)}\hat\sigma  \equiv 
\frac{d\hat\sigma^{a^\uparrow b^\uparrow\to c d}}{d\hat t} -
\frac{d\hat\sigma^{a^\uparrow b^\downarrow\to c d}}{d\hat t}
\eeq
and the chiral-odd transversity distribution (known also as $h_1$, or  
$\delta q$) is 
\beq
\label{h1}
\Delta_T f_{q/p}(x) = f_{q^\uparrow/p^\uparrow}(x) 
	 - f_{q^\downarrow/p^\uparrow}(x)\,.
\eeq
The second line in Eq.~(\ref{pi}) corresponds to the so-called Sivers 
effect, the third  to  Collins effect and the last one to the
mechanism proposed by Boer~\cite{dan2}.

\nobreak
Analogously for the  
transverse $\Lambda$ polarization in 
$pp\to \Lambda^\uparrow X$ we have   
\barr
\label{lambda}
\lefteqn{d\sigma^{\rm unp} P_\Lambda =  
	\sum_{abcd}\int \frac{dx_a dx_b}{\pi z}}\\
&& \times  
	\Big\{  \int d^2\bfk_{\perp c}\, f_{a/p}(x_a)\, f_{b/p}(x_b)\,
	\frac{d\hat\sigma}{d\hat t} (x_a,x_b;\bfk_{\perp c})\,
	 \Delta^N D_{\Lambda^\uparrow/c}(z,\bfk_{\perp c}) \nonumber\\
&& + 
	\int d^2\bfk_{\perp a}\,\Delta^Nf_{a^\uparrow/p}(x_a,\bfk_{\perp a})
	\, f_{b/p}(x_b)\, \Delta_{NN}^{(ac)}
	\hat\sigma(x_a,x_b;\bfk_{\perp a})\,
	\Delta_T D_{\Lambda/c}(z) 
	\nonumber\\
&& + 
	\int d^2\bfk_{\perp b} \,f_{a/p}(x_a)\,
	\Delta^Nf_{b^\uparrow/p}(x_b,\bfk_{\perp b})\,
	\Delta_{NN}^{(bc)}\hat\sigma(x_a,x_b;\bfk_{\perp b})\, 
	\Delta_T D_{\Lambda/c}(z) \nonumber 
\Big\}\,.
\earr
Here $\Delta_T D_{\Lambda/c}(z)$ is the analogous of $\Delta_T f_{q/p}(x)$ 
(see Eq.~(\ref{h1})). 

\vspace*{30pt}
\noindent
{\bf Phenomenology}
\vspace*{10pt}

Before entering into details of fitting procedures and
parameterizations of the relevant T-odd functions,
we try  to give a reasonable, even if qualitative, explanation of
how the intrinsic  $\bfk_\perp$ can play a crucial role in 
the observed spin hadronic asymmetries. We will refer to $\pup p\to \pi X$
(see Eq.~(\ref{pi})).

\begin{figure}[t]
\begin{center}
  \epsfig{file=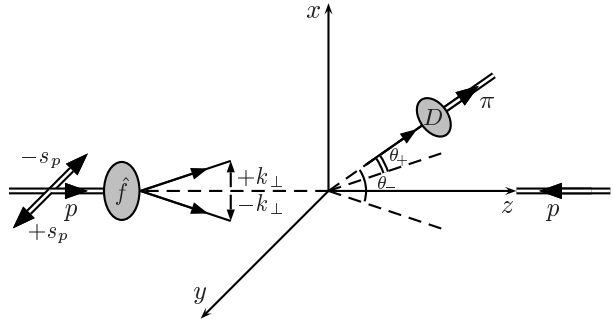,angle=+90,width=8.5cm} 
\end{center}
\caption{Pictorial representation of Sivers effect. See text for more details.
\label{ktproc}
}
\end{figure}

\begin{figure}[t]
\begin{center}
  \epsfig{file=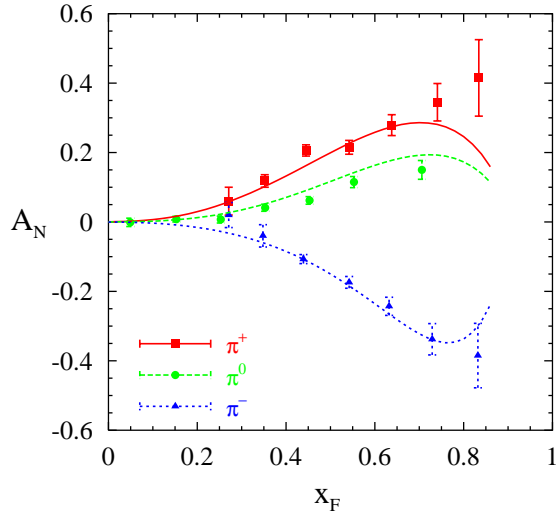,angle=-90,width=7.5cm} 
\end{center}
\caption{Best fit to $A_N$ in $p^\uparrow p\to \pi^{\pm,0}X$ as a
function of $x_F$ at $\sqrt s = 20$ GeV and $p_T=1.5$ GeV$/c$ 
(Collins effect alone). 
Data are from Ref.~\protect\cite{e704}.
\label{fitcoll}
}
\end{figure}

Let us consider  Sivers effect alone 
(similar reasonings can be done for the other contributions),  
omitting all the unnecessary variable dependences and non relevant factors.
From Eq.~(\ref{sin}) we see that 
$|\Delta^N f_{q/p^\uparrow}(x,\bfk_{\perp})|$
 reaches its maximum value at $\alpha=\pm \pi/2$. 
If we fix the scattering plane as
the $x-z$ plane with the incoming polarized proton moving along
$+\hat{\bfz}$, its $\uparrow$ 
transverse polarization is along $+\hat{\bfy}$. 
This means that $\alpha=\pm \pi/2$ 
corresponds to $\bfk_\perp$
along $\pm\hat{\bfx}$ (see Fig.~\ref{ktproc}). 
We therefore expect that at large, positive
$x_F$ ($\simorder 0.3$) 
the dominant contribution to the $\uparrow$ ($\downarrow$) 
polarized cross sections, assuming that only 
valence partons are relevant ($x_a>x_F$), is given by 
\barr
d\sigma^\uparrow  
&\sim &
	\sum_{q=u,d}
	[\hat f_{q/p^\uparrow}(+k_\perp)\,d\hat\sigma(+k_\perp) 
	+\hat f_{q/p^\uparrow}(-k_\perp)\,d\hat\sigma(-k_\perp)]D_{\pi/q}
 	\nonumber\\
& = & 
	\sum_{q=u,d}
	[\hat f_{q/p^\uparrow}(+k_\perp)\,d\hat\sigma(+k_\perp) 
	+\hat f_{q/p^\downarrow}(+k_\perp)\,d\hat\sigma(-k_\perp)]D_{\pi/q}\\
d\sigma^\downarrow 
&\sim &
	\sum_{q=u,d}
	[\hat f_{q/p^\downarrow}(+k_\perp)\,d\hat\sigma(+k_\perp)
 	+\hat f_{q/p^\downarrow}(-k_\perp)\,d\hat\sigma(-k_\perp)]D_{\pi/q}
	\nonumber\\
& = & 
	\sum_{q=u,d}
	[\hat f_{q/p^\downarrow}(+k_\perp)\,d\hat\sigma(+k_\perp)
 	+\hat f_{q/p^\uparrow}(+k_\perp)\,d\hat\sigma(-k_\perp)]D_{\pi/q}\,.
\earr
Collecting together the last expressions we have for $\pi^+$ and
$\pi^-$ production
\barr
A_N(\pi^+) & \sim & 
	[\hat f_{u/p^\uparrow}(+k_\perp) - \hat f_{u/p^\downarrow}(+k_\perp)]
	[d\hat\sigma(+k_\perp) - d\hat\sigma(-k_\perp)] \,
	D_{\pi^+/u}\\
A_N(\pi^-) & \sim & 
	[\hat f_{d/p^\uparrow}(+k_\perp) - \hat f_{d/p^\downarrow}(+k_\perp)]
	[d\hat\sigma(+k_\perp) - d\hat\sigma(-k_\perp)] \,
	D_{\pi^-/d}\,.
\earr
For the unpolarized partonic cross section,   
$d\hat\sigma(+k_\perp) > d\hat\sigma(-k_\perp) $ (see Fig.~\ref{ktproc},
and notice that $\theta_+<\theta_-$), 
therefore in order to have  $A_N(\pi^+)>0$ and $A_N(\pi^-)<0$ at
$x_F>0$ (see Fig.~\ref{fitcoll}), 
we expect 
\barr
\hat f_{u/p^\uparrow}(+k_\perp) > \hat f_{u/p^\downarrow}(+k_\perp) 
	&\Longrightarrow &
	\Delta^N f_{u/p^\uparrow}(+k_\perp)>0\\
\hat f_{d/p^\uparrow}(+k_\perp) < \hat f_{d/p^\downarrow}(+k_\perp)
	&\Longrightarrow &
	\Delta^N f_{d/p^\uparrow}(+k_\perp)<0\,.
\earr

All this 
can help also to understand why $A_N(\pi^0)$ is not 
zero. In fact  in this case, both up and down quarks enter, but 
since in the proton there are more 
up than down quarks  we expect just 
a partial cancellation and still $A_N(\pi^0)>0$.  
Two points are crucial here: $i)$ $k_\perp$ dependence in $d\hat\sigma$
($k_\perp =0 \Longrightarrow A_N=0$); $ii)$ spin and $k_\perp$ correlations 
in $\Delta^N f_{q/p^\uparrow}$, which induce an azimuthal asymmetry in
the partonic probability distributions ($\Longrightarrow 
\Delta^N f_{q/p^\uparrow}\neq 0$). 

\vspace*{30pt}
\noindent
{\bf Single spin asymmetry in {\mbox{\boldmath $p^\uparrow\!\! p\to\pi X$}}}
\vspace*{10pt}

A phenomenological study for this asymmetry 
has been carried out in a series of papers~\cite{abmsiv,abmcol}. 
For $x_F>0$ and keeping valence contributions one can
expect that the last term in Eq.~(\ref{pi}) is not important; 
on the other hand in principle both Sivers and Collins effects can play a role
simultaneously. However the  phenomenological studies presently
available   
take into account only one of these contributions at a time. 

A fit to pion data~\cite{e704} assuming Sivers effect alone and with simple
parameterizations of $\Delta^N f_{q/p^\uparrow}(x,k^0_\perp)$ in the form 
$N_q x^{a_q}(1-x)^{b_q}$, where $k^0_\perp$ is some average value of
$\bfk_\perp$ (see Ref.~\cite{abmsiv} for more details) gives reasonable
results.
The corresponding  Sivers functions  are acceptable, in particular they
satisfy the positivity condition $|\Delta^N f_{q/p^\uparrow}|\leq
2f_{q/p}$ with opposite sign for up and down contributions, as expected
from transverse momentum conservation (see also comments above).

Analogously it has been shown that Collins effect alone could explain the
data: 
a similar fitting and parameterization procedure has been adopted in
this case~\cite{abmcol}
and the quality of the fit (see Fig. \ref{fitcoll}) is 
comparable to that obtained
using only Sivers effect. However some comments are in order. 
The resulting Collins function has to saturate at large $z$ the positivity
constraint $|\Delta^N D_{\pi/q^\uparrow}|\leq 2D_{\pi/q}$.
The fitted  transversity distribution $\Delta_T f_{q/p}(x)$ 
violates the Soffer's bound.
Another fit~\cite{elliot} which 
preserves this bound gives a Collins function that almost ($\sim$ 90\%)
saturates over all $z$ values and a very small transversity
distribution.

Recently a similar study has been devoted to SSA in semi-inclusive
DIS~\cite{amcol}. 
If confirmed, these data~\cite{hermes-smc} 
indicate a large value of the Collins
function, which might then play a significant role in other processes.

\vspace*{30pt}
\noindent
{\bf Transverse {\mbox{\boldmath $\Lambda$}} polarization in 
{\mbox{\boldmath $p p\to \Lup\!\! X$}}}
\vspace*{10pt}

A huge amount of data on hyperon polarization in 
unpolarized $p-p$, $p-A$ collisions are available~\cite{data}
but no convincing
theoretical model~\cite{theo1} 
can explain them. The main features of these data, collected at $x_F>0$,  
can be summarized as follows: the transverse (with respect to the
scattering plane) $\Lambda$ polarization is negative and can be as
large as 30\% in size. $|P_\Lambda|$ grows from zero as $p_T$
increases, up to $p_T\sim 1$ GeV$/c$. At larger $p_T$ it seems to show a
plateau behaviour, up to the highest reachable $p_T$ values. The value of
$|P_\Lambda|$ in the plateau region increases almost linearly with
$x_F$. On the contrary,  $P_{\bar\Lambda}$ seems compatible with zero.

The first analysis of $P_{\Lambda,\bar\Lambda}$ within this 
formalism~\cite{lampap}  has been
recently carried out: in particular the role played by 
the {\em polarizing fragmentation function} 
$\Delta^N D_{\Lup/q}$ (second line in Eq.~(\ref{lambda})) has been
investigated. 
Indeed  the term in the last line is expected to 
give contributions at $x_F<0$.
Moreover there is some experimental evidence that the mechanism
responsible for hyperon polarization should be in the hadronization
process. This assumption can be tested looking at $P_\Lambda$ in
semi-inclusive DIS~\cite{lamprep}.

\begin{figure}[t]
\begin{center}
  \epsfig{file=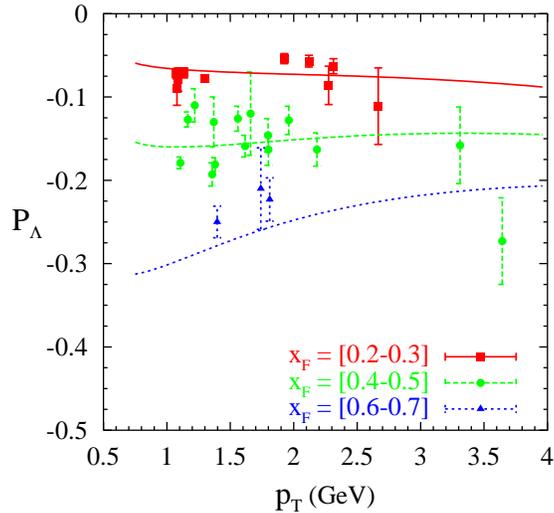,angle=-90,width=7.5cm} 
\end{center}
\caption{Best fit to $P_\Lambda$ in $p\, Be\to \Lup X$ 
as a function of $p_T$ at $\sqrt s = 80$ GeV.  
A partial collection of data from Ref.~\protect\cite{data} is shown. 
\label{fitplam}
}
\end{figure}

A fit on $P_\Lambda$ and $P_{\bar\Lambda}$ for $p_T> 1$ GeV$/c$ 
has been performed~\cite{lampap}, assuming
 simple parameterizations for $\Delta^N D_{\Lup/q}$. 
With a reasonable set of parameters 
we are able to reproduce all the
main features of data (see Fig.~\ref{fitplam}). In particular, it results 
$\Delta^N D_{\Lup/u,d} <0$, $\Delta^N D_{\Lup/s} >0$ (this can
help to explain the different behaviour of $P_\Lambda$ and 
$P_{\bar\Lambda}$) and 
$|\Delta^N D_{\Lambda/u,d}| < \Delta^N D_{\Lambda/s}$, independently
of the set of unpolarized fragmentation functions adopted.

\vspace*{30pt}
\noindent
{\bf Conclusions}
\vspace*{10pt}

We have presented a unified and consistent formalism, derived 
in the framework of pQCD, 
by extending the usual factorization theorems and including a
new class of T-odd, twist-two, spin and $\bfk_\perp$ dependent
distribution and fragmentation functions.
We have shown as it allows to describe and explain the amount of data on
single transverse  spin asymmetries observed in hadronic reactions at
moderately large $p_T$. 
A combined theoretical and experimental
analysis of several processes will allow to improve our knowledge on
such phenomena and test more deeply this formalism. 

\vspace*{30pt}
\noindent
{\bf Acknowledgments}
\vspace*{10pt}

This contribution is based on a series of papers by M. Anselmino,
D. Boer, M. Boglione and the authors. 
We acknowledge the support by CO\-FI\-NAN\-ZIA\-MENTO MURST-PRIN.

\end{document}